\documentclass[9pt,twocolumn,twoside]{arxiv}
\journal{arxiv} 
\setboolean{shortarticle}{false}

\usepackage{amsmath,graphicx}
\usepackage{bm}
\usepackage{mathtools} 
\usepackage{physics} 
\usepackage[textsize=tiny]{todonotes}
\usepackage{hyperref}
\usepackage[capitalize]{cleveref}
\usepackage{soul,xcolor}
\usepackage{multirow}
\usepackage{booktabs}

\usepackage{lineno} 
\allowdisplaybreaks[3] 
\usepackage[justification=justified]{caption} 

\title{Single-molecule orientation localization microscopy II: a performance comparison}

\author[1]{Oumeng Zhang}
\author[1,2,3,*]{Matthew D. Lew}

\affil[1]{Department of Electrical and Systems Engineering, Washington University in St. Louis, MO 63130 USA}
\affil[2]{Center for the Science and Engineering of Living Systems, Washington University in St. Louis, MO 63130 USA}
\affil[3]{Institute of Materials Science and Engineering, Washington University in St. Louis, MO 63130 USA}
\affil[*]{mdlew@wustl.edu} 

\begin{abstract}
Various techniques have been developed to measure the 2D and 3D positions and 2D and 3D orientations of fluorescent molecules with improved precision over standard epifluorescence microscopes. Due to the challenging signal-to-background ratio in typical single-molecule experiments, it is essential to choose an imaging system optimized for the specific target sample. In this work, we compare the performance of multiple state-of-the-art and commonly used methods for orientation localization microscopy against the fundamental limits of measurement precision. Our analysis reveals optimal imaging methods for various experiment conditions and sample geometries. Interestingly, simple modifications to the standard fluorescence microscope exhibit superior performance in many imaging scenarios.
\end{abstract}

\setboolean{displaycopyright}{true}

\begin{document}

\maketitle
\nolinenumbers
\section{Introduction}
\label{sec:intro}

Super-resolved, single-molecule localization microscopy (SMLM) \cite{Betzig2006,Hess2006,Rust2006,Sharonov2006} is a versatile and powerful tool for a variety of biological applications, including tracking and imaging within whole cells \cite{Huang2016,Zhu2017,Gustavsson2018}, measuring the dynamic movements of molecular motors \cite{Sosa2001,Forkey2003,Beausang2013,lippert2017angular}, and visualizing DNA conformations \cite{Ha1996,Ha1998,backer2016enhanced,Backer2019,Mazidi2019}. Not content to stop at the development of 3D SMLM \cite{VonDiezmann2017,Nehme2020}, microscopists continue to innovate spectroscopic imaging, where single-molecule (SM) fluorescence spectra \cite{Barbara2005,Zhang2015,Dong2016,Moon2017,Lee2018,Hershko2019} and SM orientations and wobbles \cite{backlund2014,ValadesCruz2016,Shaban2017,Ding2020,Lu2020} can be measured simultaneously with SM positions. New techniques also seek to attain performance approaching fundamental classical and quantum limits \cite{lupo2016,Rehacek:17,backlund2018,prasad2019,tsang2019,zhang2019fundamental,zhang2020quantum}. 

In any practical experiment, one must choose an imaging method based upon the expected signal-to-background ratio (SBR); thickness, depth, and refractive index of the sample; availability of polarization optics or phase masks; and perhaps most importantly the scientific question and measurement task at hand. An evidence-based decision hinges upon quantitative models of how spatioangular information from fluorescent SMs is mapped to photon distributions in the image plane \cite{backer2014extending,backer2015determining,Stallinga2015,Chandler:19,Chandler2019,Chandler2020}. Combining these models with classical and quantum estimation theory, one can evaluate and compare imaging techniques using the estimator-independent Cram\'er-Rao bound (CRB) \cite{moon2000mathematical} to find the optimal method for a particular application. 

In the first paper \cite{location-orientation_1} of this series, we developed fundamental precision bounds for estimating the positions and orientations of fluorescent molecules in 2D and 3D. Here, we compare various state-of-the-art and commonly used methods in 3D SM orientation localization microscopy (SMOLM) to the best-possible theoretical performance limits. We evaluate the combined orientation-position measurement precision of multiple techniques for various SBRs associated with imaging fluorescent molecules and nanoparticles. Our analysis enables scientists to choose the optimal method for various imaging scenarios, e.g., a thin 2D or thick 3D target of interest labeled with molecules rotating in either two or three dimensions. Interestingly, augmenting the standard epifluorescence microscope with simple polarizing elements produces superior performance compared to engineered point spread functions (PSFs) in a variety of scenarios.

\section{Image formation and position and orientation measurement performance}
\label{sec:model}

\begin{figure*}[ht]
    \centering
    \includegraphics[width=18.4cm]{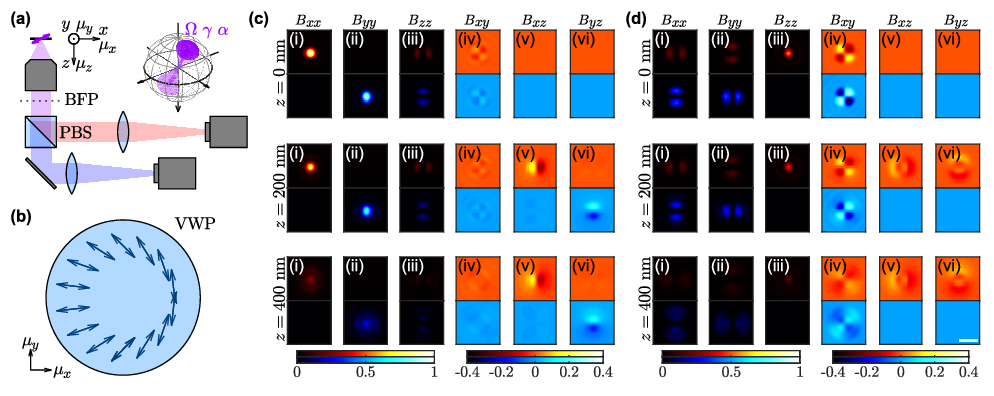}
    \caption{(a) Schematic of polarization-sensitive imaging of a fluorescent dipole emitter with orientation $\bm{\mu}=[\mu_x,\mu_y,\mu_z]^\dagger$, where the orientation axes $(\mu_x,\mu_y,\mu_z)$ are parallel with the position axes $(x,y,z)$. A polarizing beam splitter (PBS) is used to separate (blue)~$x$- and (red)~$y$-polarized light. Various engineered PSFs can be created by adding optical components, e.g., phase masks, to the back focal plane (BFP). We may parameterize rotational diffusion, or ``wobble'', using a rotational constraint factor $\gamma$, cone half angle $\alpha$, or cone solid angle $\Omega$ (inset), assuming that the molecule is uniformly diffusing within the cone. (b) A vortex (half) waveplate (VWP) can be placed at the BFP to convert radially and azimuthally polarized light to $x$- and $y$-polarized light. Arrows represent the fast axis direction of the waveplate. Basis images of (c) the $x$- and $y$-polarized standard PSF and (d) the radially and azimuthally polarized standard PSF for molecules located at (top)~$z=0$~nm, (middle)~$z=200$~nm, and (bottom)~$z=400$~nm. (i)-(vi) Basis images (i)~$B_{xx}$, (ii)~$B_{yy}$, (iii)~$B_{zz}$, (iv)~$B_{xy}$, (v)~$B_{xz}$, and (vi)~$B_{yz}$, respectively. Red: $x$-polarized image. Blue: $y$-polarized image. Colorbar: normalized intensity. Scale bar: 500~nm.}
    \label{fig:1}
\end{figure*}

\begin{figure*}[ht]
    \centering
    \includegraphics[width=18.4cm]{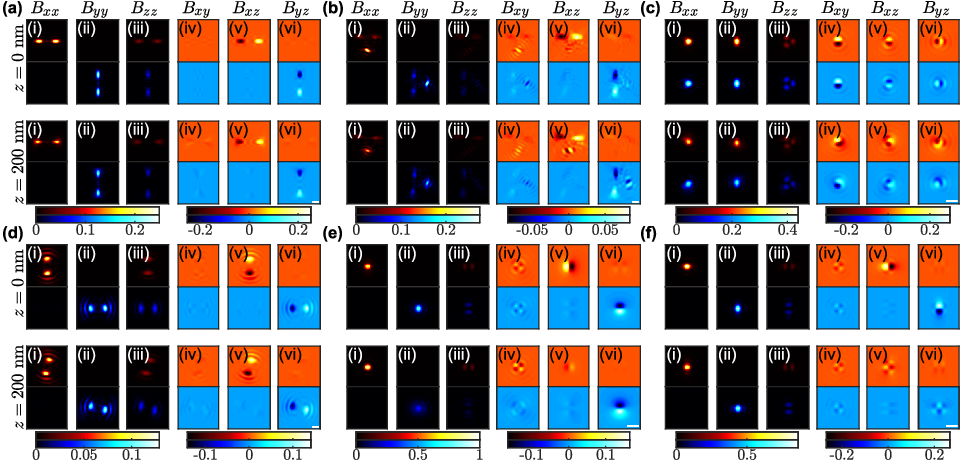}
    \caption{Basis images of (a) the Bisected PSF, (b) the Tri-spot (TS) PSF, (c) CHIDO, (d) the double-helix (DH) PSF, (e) bi(focal) plane imaging, and (f)~the astigmatic PSF for molecules located at $z=0$~nm and $z=200$~nm. (i)-(vi) Basis images (i)~$B_{xx}$, (ii)~$B_{yy}$, (iii)~$B_{zz}$, (iv)~$B_{xy}$, (v)~$B_{xz}$, and (vi)~$B_{yz}$, respectively. Red: $x$-polarized image. Blue: $y$-polarized image. We assume a 90-degree rotation of the (a)~bisected, (b) TS, (d) DH, and (f) astigmatic PSFs between $x$- and $y$-polarized imaging channels, which maximizes measurement precision. Colorbar: normalized intensity. Scale bar: 500~nm. }
    \label{fig:2}
\end{figure*}

Here, we briefly review the forward imaging model of a microscope's response to the position and orientation of single molecules; more details can be found in Ref.\ \cite{location-orientation_1}. A fluorescent molecule is modeled as an oscillating dipole with orientation represented by unit vector $\bm{\mu}=[\mu_x,\mu_y,\mu_z]^\dagger$. The second-order orientational moments of $\bm{\mu}$ are given by 
\begin{equation}
    m_{ij}=\frac{1}{T}\int_0^T\mu_i\mu_j\,dt,
\end{equation}
where $\{i,j\}\in\{x,y,z\}$. For molecules symmetrically wobbling around an average orientation $\bar{\bm{\mu}}=[\bar{\mu}_x,\bar{\mu}_y,\bar{\mu}_z]^\dagger$ with rotational constraint $\gamma$ [\cref{fig:1}(a), \cite{zhang2019fundamental}], the second moments characterize both average orientation and wobble via
\begin{linenomath}\begin{subequations}
\begin{align}
    m_{ii} &= \gamma\bar{\mu}_i^2+\frac{1-\gamma}{3} \\
    m_{ij} &= \gamma\bar{\mu}_i\bar{\mu}_j & i,j\in\{x,y,z\},\,i\neq j.
\end{align}
\end{subequations}\end{linenomath}
For a translationally fixed molecule located at position $\bm{r}=[x,y,z]^\dagger$, the fluorescence intensity relayed by an imaging system is given by
\begin{linenomath}\begin{multline}
    I(\xi,\eta;\bm{r},\bm{m})=\left[B_{xx}(\xi,\eta;\bm{r}),B_{yy}(\xi,\eta;\bm{r}),B_{zz}(\xi,\eta;\bm{r}),\right.\\
    \left.B_{xy}(\xi,\eta;\bm{r}),B_{xz}(\xi,\eta;\bm{r}),B_{yz}(\xi,\eta;\bm{r})\right]\,\bm{m},
\end{multline}\end{linenomath}
where $(\xi,\eta)$ represents the image plane coordinate system and $\bm{m}=[m_{xx},m_{yy},m_{zz},m_{xy},m_{xz},m_{yz}]^\dagger$. The basis images of the imaging system $B_{ij}(\xi,\eta;\bm{r})$ are given by
\begin{linenomath}\begin{subequations}
\begin{align}
    B_{ii} &= |G_i|^2           & i\in\{x,y,z\}\\
    B_{ij} &= G_iG_j^*+G_i^*G_j & i\in\{x,y,z\},\,i\neq j,
\end{align}
\end{subequations}\end{linenomath}
where $G_i(\xi,\eta;\bm{r})$ are the basis fields corresponding to fixed dipoles with orientations $\mu_i=1$ \cite{location-orientation_1}. In this work, we model a polarization-sensitive imaging system [\cref{fig:1}(a)] created by placing a polarizing beamsplitter in the fluorescence detection path. The basis images $B_{ii}$ and $B_{ij}$ therefore represent $x$-polarized [red in \cref{fig:1}(c,d) and \cref{fig:2}] and $y$-polarized [blue in \cref{fig:1}(c,d) and \cref{fig:2}] images detected simultaneously. Similar to the standard imaging system PSF, shift invariance also holds for these basis images. Throughout this paper, we use a fluorescence wavelength in air of 600~nm, an objective lens numerical aperture (NA) of 1.4, and a lens immersion medium of refractive index  $n=1.515$.

One commonly used method to measure molecular orientation is separating the $x$- and $y$-polarized emission light [$xy$Pol, \cref{fig:1}(c), \cite{mortensen2010optimized}], which can be used to distinguish $x$- and $y$-oriented molecules with high precision. However, its sensitivity to distinguish molecular orientation $[\mu_x,\mu_y,\mu_z]$ from $[-\mu_x,\mu_y,\mu_z]$ is low due to the weak basis image $B_{xy}$ [\cref{fig:1}(c)(iv)]. To overcome this limitation, we propose a method where a vortex (half) waveplate [VWP, \cref{fig:1}(b)] is placed at the back focal plane (BFP) of the imaging system to convert radially  and azimuthally polarized light to $x$- and $y$- polarized light, i.e., separating radially from azimuthally polarized fluorescence [raPol, \cref{fig:1}(d), \cite{lew2014azimuthal,backlund2016removing,zhang2020quantum}]. This strategy is optimal for estimating all in-($xy$)plane second-order orientational moments $[m_{xx},m_{yy},m_{xy}]$ \cite{location-orientation_1}. 

In our previous work \cite{location-orientation_1}, we found that both $xy$Pol and raPol lack the sensitivity to measure the out-of-plane moments $m_{xz}$ and $m_{yz}$ due to the extremely weak basis images $B_{xz}$ and $B_{yz}$ [\cref{fig:1}(c,d)(v,vi)]. Either phase or polarization modulation of the light at the BFP is required to improve the precision in measuring these moments. Besides PSF engineering, we note that simple defocusing of a sample adds a complex phase modulation to the optical field at the BFP, causing the basis images $B_{xz}$ [\cref{fig:1}(c,d)(v)] and $B_{yz}$ [\cref{fig:1}(c,d)(vi)] to contain much stronger contrast. Further, we notice that when the molecule is defocused by $z=200$~nm, the resulting PSFs expand to only a small degree compared to their in-focus counterparts. This observation implies that orientation measurement precision can be greatly improved by tolerating a minor degradation in lateral localization precision. 

Here, we evaluate the limit of measurement precision for a given technique by computing the standardized generalized variance (SGV) \cite{sengupta1987tests} associated with the Fisher information (FI) matrices 1) $\bm{\mathcal{J}_m}$ for estimating the orientational second moments $\bm{m}$ and 2) $\bm{\mathcal{J}_r}$ for estimating the position $\bm{r}$ of an SM, thereby summarizing the multiparameter CRB. In the following sections of this paper, we proceed with investigating overall orientation-position estimation precision and present a comprehensive performance comparison between the polarized standard PSFs and several state-of-the-art and commonly used techniques for 3D orientation measurements and 3D localization (\cref{fig:2}). The Bisected [BS, \cref{fig:2}(a), \cite{backer2014bisected}] and Tri-spot [TS, \cref{fig:2}(b), \cite{zhang2018imaging}] PSFs are designed to generate a multi-spot image for each single molecule by placing various linear phase ramps in various sections of the BFP. For these PSFs, axial position is encoded in the distance between the spots, and orientation information is encoded in the relative brightness between the spots. Coordinate and Height super-resolution Imaging with Dithering and Orientation (CHIDO) [\cref{fig:2}(b), \cite{curcio2019birefringent}] uses a stressed-engineered optic (SEO) to modulate phase and polarization, thereby creating a set of linearly independent basis images that rotate when a molecule is defocused. The precise features of this PSF depend on a stress coefficient, which we set to $0.5\pi$ such that CHIDO performance is optimized for the scenarios studied here. This coefficient is different from that used in \cite{curcio2019birefringent,location-orientation_1} ($1.2\pi$). The double-helix [DH, \cref{fig:2}(d), \cite{Pavani2008,Pavani2009}] PSF is designed such that two spots revolve around another as a function of the axial position $z$ of a SM; orientation information is also encoded in the relative brightness of the spots \cite{Backlund2012}. We also include two simple polarized variations of the standard PSF, bi(focal) plane imaging [\cref{fig:2}(e), focal planes located at $z=\pm175$~nm \cite{Kao1994,Agrawal2012}] and the astigmatic PSF [\cref{fig:2}(f), 200~nm between focal planes with the circle of least confusion at $z=200$~nm \cite{Toprak2007}].

\section{Optimal orientation localization microscopy approaching the quantum limit}

\begin{figure}[ht]
    \centering
    \includegraphics[width=8.8cm]{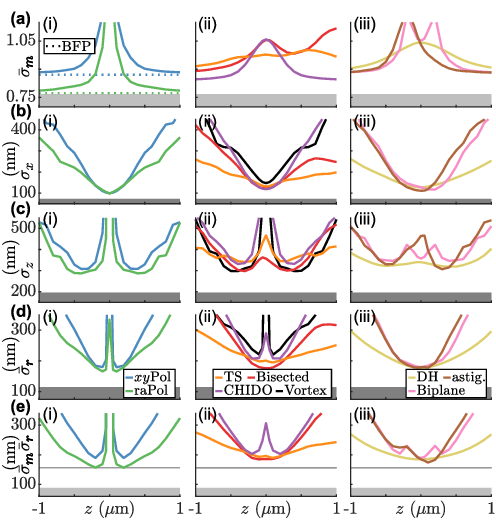}
    \caption{Limit of precision for measuring (a) 3D molecular orientation $\bm{m}$, (b) lateral position $x$, (c) axial position $z$, (d) overall 3D position $\bm{r}$, and (e) combined 3D orientation and position of freely rotating molecules (i.e., $\bm{m}=[1,1,1,0,0,0]^\dagger/3$) with one photon detected using (i) polarized standard PSFs, (ii) PSFs designed for orientation measurements, and (iii) PSFs designed for 3D localization. Dotted line represents the precision when the detectors are placed at the BFP instead of the image plane; gray line represents the best overall orientation localization performance, attained by the raPol PSF at a defocus of \textasciitilde200~nm. Blue: standard PSF with $x$ and $y$-polarization separation ($xy$Pol), green: standard PSF with radial and azimuthal polarization separation (raPol), red: Bisected PSF, orange: Tri-spot (TS) PSF, purple: CHIDO, black: Vortex PSF, yellow: double-helix (DH) PSF, pink: biplane imaging, brown: astigmatic (astig.) PSF. Gray areas are bounded from above by the (a) classical and (b-d) quantum limits derived in \cite{location-orientation_1}.}
    \label{fig:3}
\end{figure}

We begin by evaluating measurement performance for an isotropic emitter, i.e., $\bm{m}=[1,1,1,0,0,0]^\dagger/3$. First, we evaluate the overall limit of precision for measuring the six 3D orientational second moments [\cref{fig:3}(a)], given by 
\begin{equation}
    \bar{\sigma}_{\bm{m}}=\left[\det(\bm{\mathcal{J}_m})^{1/6}\right]^{-1/2},
\end{equation}
the square-root of the SGV, where $\det(\cdot)$ denotes the matrix determinant. Traditional performance metrics, e.g., the average CRB, generally exhibit similar trends as the SGV. However, SGV also accounts for correlations in measurement sensitivity between parameters \cite{location-orientation_1}. For all methods we evaluated except TS, which is designed specifically to measure intensity in each region of the BFP regardless of defocus-induced phase variations, orientation precision improves when the emitter is shifted away from the focal plane(s). Interestingly, if we directly image the radially and azimuthally polarized intensity distribution in the BFP [\cref{fig:3}(a)(i)], we can measure the second moments with precision close to (within 1\% of) the fundamental bound \cite{location-orientation_1}; however, such an approach can only measure one emitter at a time. Moving to the image plane, we find that a defocused raPol PSF outperforms all other image plane techniques when defocused. Here, we also include a recently developed method, the Vortex PSF \cite{Hulleman2020.10.01.322834}, which adds a vortex phase plate, commonly used in STED nanoscopy, in the detection path of a single-channel imaging system. However, we notice that its orientation estimation precision suffers due to its lack of polarization sensitivity; its best-possible precision lies above the plotting ranges in \cref{fig:3}. Therefore, we do not include it in further analysis.

Next, we evaluate the best-possible precision of localizing emitters in 3D space $\bm{r}=[x,y,z]^\dagger$. The lateral precision [\cref{fig:3}(b)] of all PSFs degrades when the emitter is defocused due to PSF expansion and the resulting reduction of peak SBR. However, axial precisions [\cref{fig:3}(c)] improve when the molecule is slightly defocused by several hundred nanometers, except for the DH PSF \cite{Pavani2009}, which is specifically designed to achieve a more uniform $z$ precision. Turning to the overall 3D localization precision [\cref{fig:3}(d)] given by 
\begin{equation}
    \bar{\sigma}_{\bm{r}}=\left[\det(\bm{\mathcal{J}}_{\bm{r}})^{1/3}\right]^{-1/2},
\end{equation} 
we find that the polarized standard PSFs and CHIDO perform best when defocus is approximately 100-200~nm. For the Bisected PSF, TS PSF, astigmatic PSF \cite{Kao1994}, and biplane imaging \cite{Toprak2007}, the overall precision is relatively uniform within the -200 to 200~nm defocus range.

We next define the limit of overall orientation-position precision as $\bar{\sigma}_{\bm{m}}\bar{\sigma}_{\bm{r}}$ [\cref{fig:3}(e)], which can be viewed as the geometric average of the estimation precisions for the second-order orientational moments $m_{ij}$ and 3D position $\bm{r}$ combined. For the polarized standard ($xy$Pol and raPol) PSFs, CHIDO, biplane imaging, and the astigmatic PSF, our computations show that overall precision is best when the molecule is defocused by 100 to 200~nm away from the focal plane(s), which matches our intuition in \cref{sec:model}. Notably, the raPol PSF, at a defocus of \textasciitilde200~nm, exhibits the best combined orientation-position precision compared to all other methods.

\begin{figure}[ht!]
    \centering
    \includegraphics[width=8.8cm]{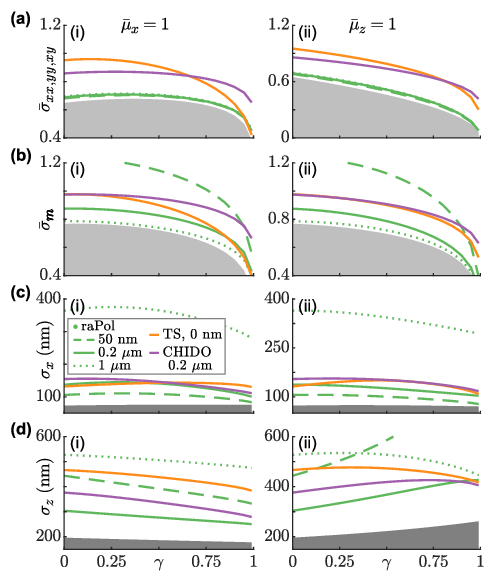}
    \caption{Limit of precision of measuring the (a) in-($xy$) plane orientation $[m_{xx},m_{yy},m_{xy}]$, (b) 3D orientation $\bm{m}$, (c) lateral position $x$, and (d) axial position $z$ of molecules wobbling around the (i)~$\mu_x$ axis and (ii)~$\mu_z$ axis with one photon detected. Green: radially and azimuthally polarized standard PSF (raPol) defocused by 50~nm (dashed), 200~nm (solid), and 1000~nm (dotted); orange: Tri-spot (TS) PSF; purple: CHIDO. Gray areas are bounded from above by the (a-b) classical and (c-d) quantum limits derived in \cite{location-orientation_1}.}
    \label{fig:4}
\end{figure}

Our performance analysis thus far has focused on isotropic emitters and freely rotating molecules. Next, we quantify measurement precision for molecules symmetrically wobbling around the $\mu_x$ and $\mu_z$ axes and evaluate performance against fundamental bounds  \cite{location-orientation_1}. We compare raPol at various values of defocus to the TS PSF and CHIDO at their optimal defocus positions. We notice that defocusing has no effect on the precision of measuring molecular orientation in the $xy$ plane [\cref{fig:4}(a)], defined as
\begin{equation}
    \bar{\sigma}_{xx,yy,xy}=\left[\det\left(\bm{\mathcal{J}}_{m_{xx},m_{yy},m_{xy}}\right)^{1/3}\right]^{-1/2}.
\end{equation}
That is, raPol's in-plane orientation precision is near the fundamental limit derived in \cite{location-orientation_1} for molecules positioned 50-1000~nm from the focal plane. The overall precision $\bar{\sigma}_{\bm{m}}$ of measuring all (3D) second moments using raPol improves as defocus increases [\cref{fig:4}(b)], indicating that defocusing improves the precision of measuring $m_{xz}$ and $m_{yz}$ without sacrificing estimation precision for the in-plane moments; defocused raPol imaging ($z=1000$~nm) performs similarly to direct BFP imaging. The orientation precision $\bar{\sigma}_{\bm{m}}$ when $z=200$~nm is 15\% worse on average compared to that at $z=1000$~nm but is still superior to that of TS and CHIDO.

In terms of localization, the lateral precision of raPol [\cref{fig:4}(c)] naturally worsens as defocus increases for molecules wobbling around both the $\mu_x$ and $\mu_z$ axes. The precision when defocus $z=200$~nm is 23\% worse on average compared to that at $z=50$~nm but is still comparable to the precision of TS and CHIDO. The axial precision using raPol [\cref{fig:4}(d)] is best when $z=200$~nm, making it more precise than TS and CHIDO for most orientations. Considering the overall orientation-position precision $\bar{\sigma}_{\bm{m}}\bar{\sigma}_{\bm{r}}$ of raPol, we observe that defocusing by 200~nm provides a good compromise; this condition has much better orientation precision compared to that at 50~nm defocus without a severe sacrifice in localization precision like defocusing by 1000~nm. In fact, our analysis shows that using raPol at a defocus of $z=200$~nm exhibits superior orientation-position measurement precision compared to all other methods.

\section{Orientation localization performance under practical imaging conditions}
\label{sec:indexMismatch}

\begin{figure}[ht!]
    \centering
    \includegraphics[width=8.8cm]{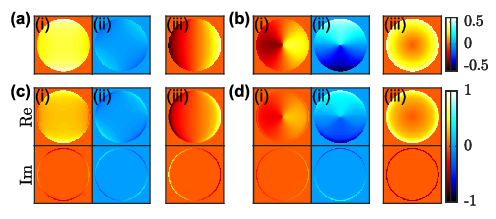}
    \caption{Basis fields $G_x$ in the (i)~$x$-polarized and (ii)~$y$-polarized channels and (iii)~$G_z$ in the $x$-polarized channel at the BFP for (a)~$x$ and $y$-polarization separation and (b) radial and azimuthal polarization separation when the sample refractive index matches the lens immersion medium (1.515) and (c-d) when the sample refractive index (1.33) is smaller than the imaging medium and numerical aperture (NA=1.4). Basis fields not shown here are either rotated versions of the ones shown here or zero.}
    \label{fig:5}
\end{figure}

In biological imaging, the refractive index (RI) of the medium surrounding the fluorescent molecules of interest is often different from the designed immersion medium of the objective lens itself. These effects must be carefully calibrated since model mismatch likely creates significant biases in the orientation-position measurement \cite{Petrov2020}. If the objective lens's numerical aperture (NA) is smaller than or equal to the sample RI or the fluorophores are far away from the RI interface, then the typical dipole emission pattern is observed at the BFP [\cref{fig:5}(a,b)]. The resulting orientation-position precisions will exhibit trends similar to those in \cref{fig:3,fig:4}. However, when the NA of the objective lens is greater than the sample RI and the fluorophores are near the RI interface, supercritical light is captured by the objective lens, resulting in very different optical fields at the BFP \cite{Ruckstuhl2000,axelrod2012fluorescence}. For molecules located at the interface between a sample with RI equal to that of water (1.33) and the objective's immersion medium, the supercritical light ring [\cref{fig:5}(c,d)] contains a non-uniform phase pattern at the BFP, thereby breaking the degeneracy for measuring the out-of-plane moments $m_{xz}$ and $m_{yz}$ for in-focus emitters. Further, defocusing in the $+z$ direction no longer produces PSFs degenerate with defocusing in the $-z$ direction. Therefore, axial localization precision for molecules close to the focal plane is also greatly improved. 

Another major factor that limits measurement precision is background photons detected during fluorescence imaging. In this section, we evaluate how mismatched RI and limited SBR affect overall instrument performance by simulating molecules embedded in a sample with an RI of 1.33 [\cref{fig:6}(a)]. We denote the distance between the molecule and the RI interface by $h$ and the position of the nominal focal plane [i.e., the focal plane position when imaging without a RI mismatch, which is above the actual focal plane, \cref{fig:6}(a)] by $z$.
We characterize performance for a typical SM SBR (1,000 signal photons and 5 background photons/pixel) and a high SBR typical of quantum nanorods (30,000 signal photons \cite{lippert2017angular} and 5 background photons/pixel); these SBRs hold for rotationally fixed molecules located at the RI interface with $\bar{\mu}_z=0$ (perpendicular to the optical axis). We choose to hold the fluorescence photon emission rate fixed for all conditions; therefore, the SBR will vary for molecules tilted away from the coverslip ($\bar{\mu}_z>0$) or positioned away from the RI interface [\cref{fig:6}(b)].

\subsection{Thin planar samples}

\begin{figure}[t!]
    \centering
    \includegraphics[width=8.8cm]{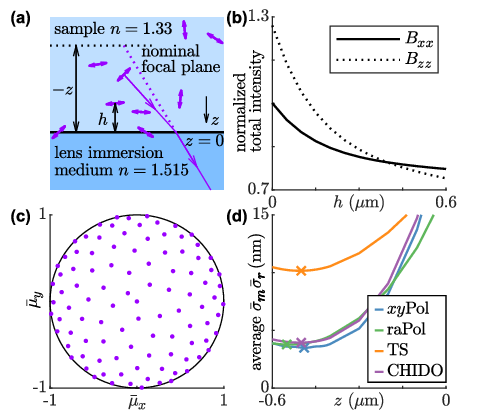}
    \caption{(a) Schematic of a sample containing molecules distributed at various heights $h$ above the refractive index (RI) interface. The nominal focal plane (dotted black line) is located at a distance $-z$ above ($+z$ below) the RI interface ($z=0$). (b)~The normalized integrated intensities of basis images $B_{xx}$ and $B_{zz}$, corresponding to fixed dipoles with $\mu_x=1$ and $\mu_z=1$, respectively, change with a molecule's height $h$ above the interface. Due to symmetry, the total intensity of $B_{xx}$ is equal to that of $B_{yy}$. (c) Two hundred mean orientations $\bar{\bm{\mu}}$ uniformly sampled on the unit sphere ($\bar{\mu}_z>0$). (d) Average orientation-position estimation precision $\bar{\sigma}_{\bm{m}}\bar{\sigma}_{\bm{r}}$ for molecules located within 600~nm of the RI interface as function of the focal plane position $z$. Crosses denote the focal plane placement that maximizes overall orientation localization precision throughout the sample. Blue: $x$- and $y$-polarized standard PSF ($xy$Pol); green: radially and azimuthally polarized standard PSF (raPol); orange: Tri-spot (TS) PSF; purple: CHIDO.}
    \label{fig:6}
\end{figure}

First, we evaluate the orientation-position precision of the aforementioned techniques for molecules located at the RI interface [$h=0$, \cref{fig:6}(a)]. However, instead of using freely rotating molecules, we sample estimation performance for molecules symmetrically wobbling around 200 average orientations $\bar{\bm{\mu}}$ uniformly sampled on the orientation unit sphere [\cref{fig:6}(c)] with rotational constraint $\gamma=0.8$ \cite{zhang2019fundamental}, which is equivalent to a cone surface area of $\Omega=0.879$ sr or a cone half angle of $\alpha = 30.7^\circ$ if the molecule is uniformly wobbling within it [\cref{fig:1}(a)]. We report measurement performance as the average estimation precision across all sampled orientations.

\begin{figure}[ht!]
    \centering
    \includegraphics[width=8.8cm]{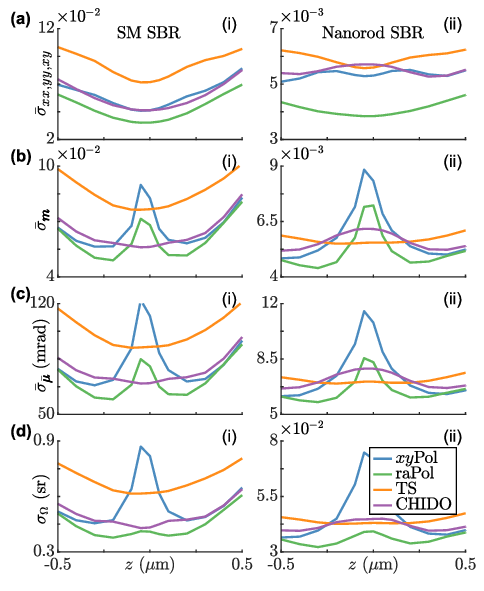}
    \caption{Limit of precision for measuring (a) the $xy$-plane second-order orientational moments $[m_{xx},m_{yy},m_{xy}]$, (b) the 3D orientation $\bm{m}$, (c) the average orientation $\bar{\bm{\mu}}$, and (d) the cone solid angle $\Omega$ of a molecule uniformly wobbling in a cone. Estimation precisions are reported for molecules lying at a water-glass interface with rotational constraint $\gamma=0.8$ and averaged over all possible mean orientations $\bar{\bm{\mu}}$ (\cref{fig:6}(c)). Two signal to background ratios (SBRs) are considered: (i) one typical of single molecules (SMs) with 1,000 signal photons and (ii) one typical of quantum nanorods with 30,000 signal photons; both use 5 background photons per $58.5\times58.5$~nm$^2$ pixel. Blue: $x$- and $y$-polarized standard PSF ($xy$Pol); green: radially and azimuthally polarized standard PSF (raPol); orange: Tri-spot (TS) PSF; purple: CHIDO.}
    \label{fig:7}
\end{figure}

In the presence of background, larger PSFs impart increasingly poor precision due to a reduced imaging SBR. The precision of measuring in-plane orientations as a function of defocus thus scales intuitively [\cref{fig:7}(a)]. The radially and azimuthally polarized PSF, which performs closely to the fundamental bound without background photons \cite{location-orientation_1}, has better precision than $xy$Pol, TS, and CHIDO. Although collecting supercritical light improves the sensitivity of measuring $m_{xz}$ and $m_{yz}$ for molecules that are in focus, the 3D orientation precision [\cref{fig:7}(b)] still improves with defocus for the polarized standard PSFs.  When the focal plane is placed 200-300~nm into the lens immersion medium, raPol exhibits the best precision among these techniques for all SBRs [\cref{fig:7}(b)]. Due to its large size, the TS PSF has comparatively worse precision for SMs [\cref{fig:7}(a,b)(i)] but has comparable or superior precision to other techniques for quantum rods [\cref{fig:7}(a,b)(ii)]. It also performs more uniformly within the $|z|<500$~nm range compared to the polarized standard PSFs and CHIDO; the TS PSF effectively has an increased depth of field over the other methods due to its partitioning of the BFP. The peak-to-valley difference in precision is 47\% of the average precision, which is much smaller than 90\% for the polarized standard PSFs and 77\% for CHIDO. Interestingly, CHIDO exhibits its best precision for in-focus emitters at SM SBRs, but for high nanorod SBRs, defocus improves measurement precision [\cref{fig:7}]; more photons enable fine features of the defocused CHIDO PSF to be utilized for improved sensitivity.

To provide physical intuition for these precision limits, we also evaluate the precision of estimating parameters of the uniform ``wobble in a cone'' model: the average orientation $\bar{\bm{\mu}}$ of the SM and the solid angle $\Omega=(3-\sqrt{8\gamma+1})\pi$ [\cref{fig:1}(a)] of the cone in which it diffuses. The uncertainty $\bar{\sigma}_{\bar{\bm{\mu}}}$ in mean orientation is given by
\begin{equation}
    \bar{\sigma}_{\bar{\bm{\mu}}}=\left[\bar{\mu}_z\det\left(\bm{\mathcal{J}}_{\bar{\mu}_x,\bar{\mu}_y}\right)^{1/2}\right]^{-1/2}
\end{equation}
[\cref{fig:7}(c)], which represents an arc length on the unit orientation sphere [\cref{fig:1}(a)]. This measurement precision largely scales with that of estimating all second-order orientational moments $\bar{\sigma}_{\bm{m}}$ [\cref{fig:7}(b)]; raPol has an average orientation measurement precision  of 72.2 mrad at SM SBRs and 6.7 mrad at nanorod SBRs. Interestingly, calculating the best-possible precision $\sigma_\Omega$ in wobble angle [\cref{fig:7}(d)] shows that raPol has largely uniform and superior performance over a large defocus range; raPol has an average wobble measurement precision of 0.439 sr for SMs (0.036 sr for nanorods).

\begin{figure}[ht!]
    \centering
    \includegraphics[width=8.8cm]{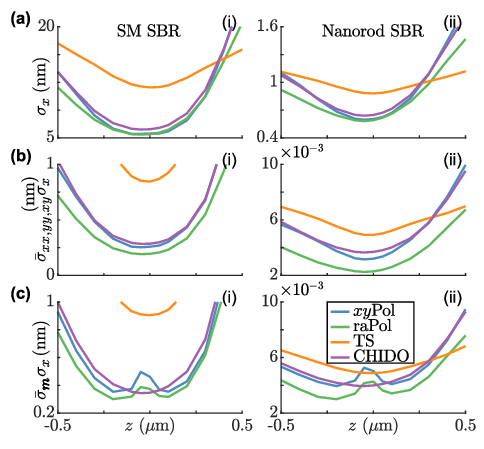}
    \caption{Limit of precision for measuring the (a) lateral position $[x,y]$, (b) 2D in-plane orientation $[m_{xx},m_{yy},m_{xy}]$ and lateral position, and (c) combined 3D orientation $\bm{m}$ and lateral position of a molecule uniformly wobbling in a cone. Estimation precisions are reported for molecules lying at a water-glass interface with rotational constraint $\gamma=0.8$ and averaged over all possible mean orientations $\bar{\bm{\mu}}$ (\cref{fig:6}(c)).  Two signal to background ratios (SBRs) are considered: (i) one typical of single molecules (SMs) with 1,000 signal photons and (ii) one typical of quantum nanorods with 30,000 signal photons; both use 5 background photons per $58.5\times58.5$~nm$^2$ pixel. Blue: $x$- and $y$-polarized standard PSF ($xy$Pol); green: radially and azimuthally polarized standard PSF (raPol); orange: Tri-spot (TS) PSF; purple: CHIDO.}
    \label{fig:8}
\end{figure}

We next explore how measurement precision scales as the imaging task becomes increasingly complex, holding the image SBR fixed. If one is only interested in localizing a molecule in 2D [\cref{fig:8}(a)], then localization precision degrades with increasing defocus $z$ as expected. The polarized standard PSFs and CHIDO are similar to one another since their sizes are comparable; both have superior precision compared to the TS PSF, especially for SM SBRs. The TS PSF exhibits a more uniform precision across the 1 $\mu$m depth range.

Measuring the in-plane position $[x,y]$ and additionally the in-plane orientational moments $[m_{xx},m_{yy},m_{xy}]$ of SMs is important when, for example, performing TAB SMOLM of amyloid fibers \cite{Ding2020}. Therefore, we use the product $\bar{\sigma}_{xx,yy,xy}\,\sigma_x$ to evaluate the overall 2D orientation-position precision [\cref{fig:8}(b)]. When the (thin) sample is located at the focal plane, raPol performs the best since it has the highest precision for both 2D position and 2D orientation measurements. 

In another scenario, molecules may be confined within or near a single $z$ plane but their orientations could lie anywhere in 3D space. This situation occurs, for example, when using PAINT to perform SMOLM of a supported lipid bilayer \cite{Lu2020}. Therefore, 3D orientation and 2D localization performance can be measured by the product of the lateral position precision $\sigma_x$ and the 3D orientation precision $\bar{\sigma}_{\bm{m}}$ [\cref{fig:8}(c)]. When the (thin) sample is located at $z=\pm200$~nm, raPol has the best overall 3D orientation localization performance, with $xy$Pol and CHIDO close behind. Overall, our analysis shows that to obtain improved orientation-position precision using the family of standard PSFs, one should modestly defocus the sample by \textasciitilde200~nm.

\subsection{Thick three-dimensional samples}

Since the intensity of supercritical light decays rapidly with increasing distance between the molecule and RI interface, orientation localization performance for a sample that is much thicker than the wavelength will yield trends similar to those shown in \cref{fig:3}(e). In this section, we analyze the precision of measuring the 3D orientation and position of molecules within a sample of thickness of $h_{\max}=600$~nm [\cref{fig:6}(a)], such that supercritical light is captured for a majority of the sample; supercritical light comprises only 3\% of the detected fluorescence from a molecule at $h=h_{\max}$ \cite{JamesShirley2018}.

First, for each technique, we find the optimal position of the objective lens focal plane by computing the average measurement precision $\bar{\sigma}_{\bm{m}}\bar{\sigma}_{\bm{r}}$ for isotropic emitters located across all depths $h$ between 0 and 600~nm at a typical SM SBR; the optimal focal plane position $z$ minimizes this average precision. We find that $xy$Pol, raPol, TS, and CHIDO achieve their best precisions when $z=-490$, $-550$, $-500$, and $-500$~nm, respectively [\cref{fig:6}(d)]. The polarized standard PSFs and CHIDO exhibit similar precision; the difference in the optimal average $\bar{\sigma}_{\bm{m}}\bar{\sigma}_{\bm{r}}$ is within 12\%. Similar to 3D orientation and 2D localization measurements [\cref{fig:8}(c)], TS performs worse compared to other methods for measuring 3D orientation and 3D position.

\begin{figure}[ht!]
    \centering
    \includegraphics[width=8.8cm]{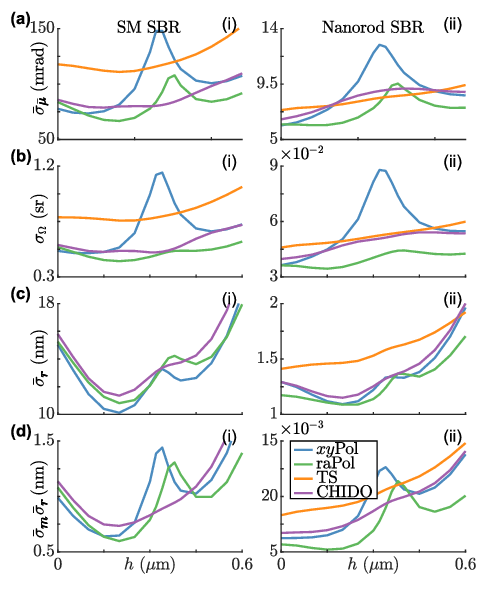}
    \caption{Limit of precision for measuring the (a) average orientation $\bar{\bm{\mu}}$, (b) cone solid angle $\Omega$, (c) 3D position $\bm{r}$, and (d) 3D position and 3D orientation of (i) a single molecule (SM) with 1,000 signal photons and (ii) a quantum nanorod with 30,000 signal photons and 5 background photons per $58.5\times58.5$~nm$^2$ pixel detected. Estimation precisions are reported for molecules with rotational constraint $\gamma=0.8$ and averaged over all possible mean orientations $\bar{\bm{\mu}}$ (\cref{fig:6}(c)). Blue: $x$- and $y$-polarized standard PSF ($xy$Pol); green: radially and azimuthally polarized standard PSF (raPol); orange: Tri-spot (TS) PSF; purple: CHIDO. For each method, the nominal focal plane was placed at the optimal $z$ (\cref{fig:6}(d)) to maximize overall performance.} 
    \label{fig:9}
\end{figure}

We now quantify the overall orientation localization precision of these techniques (using the optimal value of $z$) for thick samples by averaging measurement precision across 200 uniformly sampled mean molecular orientations and depths $h \le h_{\max}$ while holding wobble fixed ($\gamma=0.8$). The precision of measuring the 3D average orientation $\bar{\sigma}_{\bar{\bm{\mu}}}$ [\cref{fig:9}(a)] and wobble cone solid angle $\bar{\sigma}_{\Omega}$ [\cref{fig:9}(b)] exhibit similar trends with axial position $h$ compared to that for defocusing thin planar samples [\cref{fig:7}(c,d)]. On average, raPol measures average orientation $\bar{\bm{\mu}}$ 2\% and 11\% more precisely than CHIDO does under SM and nanorod SBRs, respectively; raPol has an average orientation measurement precision  of 84.5 mrad at SM SBRs and 7.6 mrad at nanorod SBRs. However, CHIDO and TS perform more uniformly over the depth range compared to the polarized standard PSFs. As expected, raPol also exhibits the best precision for measuring molecular wobble $\Omega$ at almost every molecule depth $h$; raPol has an average wobble measurement precision of 0.496~sr for SMs (0.040~sr for nanorods). Intuitively, measurement precision is better for molecules located closer to the RI interface due to the increased contribution of supercritical light. 

The average 3D localization precision $\bar{\sigma}_{\bm{r}}$ [\cref{fig:9}(c)] using CHIDO is 8\% and 4\% worse than that of $xy$Pol and raPol respectively at SM SBRs (3\% worse than $xy$Pol and 8\% worse than raPol at nanorod SBRs). On average, $xy$Pol can achieve 13.0~nm localization precision in 3D for SMs and 1.3~nm precision for nanorods, while raPol has best-possible 3D localization precisions of 10.8~nm at SM SBRs and 1.1~nm at nanorod SBRs. However, CHIDO performs more uniformly for molecules at intermediate depths $h\in[150,450]$~nm. TS measurement precision is 60\% worse than CHIDO for SM SBRs and is therefore not shown in \cref{fig:9}(c)(i). Its precision is more uniform across the entire 600-nm range due to its large depth of field [\cref{fig:9}(c)(ii)].

Examining overall 3D orientation and 3D localization measurement performance, raPol performs the best out of all techniques under both SM and nanorod SBRs over the entire 600~nm range. Its average precision $\bar{\sigma}_{\bm{m}}\bar{\sigma}_{\bm{r}}$ is 10\% and 5\% better compared to those of $xy$Pol and CHIDO at SM SBRs [\cref{fig:9}(d)(i)] and 20\% and 14\% better at nanorod SBRs [\cref{fig:9}(d)(ii)], respectively. However, since CHIDO has more uniform performance for both 3D orientation [\cref{fig:9}(a)] and position [\cref{fig:9}(c)], its overall measurement precision is also much more uniform, especially within $h\in[150,450]$~nm.

\section{Discussion and Conclusion}
\label{sec:discussion}

In the first paper in this series \cite{location-orientation_1}, we presented a mathematical framework for finding the fundamental sensitivity limits of measuring the orientations and positions of single molecules, in both 2D and 3D, using any imaging system. Here, we evaluate multiple state-of-the-art and commonly used imaging techniques and compare their performances to the fundamental bounds. We find that the radially  and azimuthally polarized standard PSF (raPol), which only requires a simple addition of a vortex (half) waveplate at the BFP of a polarization-sensitive epifluorescence microscope, achieves nearly the maximum (optimal) sensitivity attainable when measuring both lateral (2D) molecular orientation and 3D molecular orientation in thin samples (\cref{fig:3,fig:4}). However, none of the methods we evaluated performs closely to the maximum possible performance for measuring both 3D orientation and 3D position (\cref{fig:3}), which suggests that there still exists room for improving SMOLM methods and extracting the maximum information possible from each detected photon. Interferometric detection with one or multiple objective lenses \cite{Hell:94,Shtengel2009,Aquino2011,Huang2016,backlund2018,zhang2020quantum} is one possible avenue to pursue, as interferometry enables the full complex wavefunction to be measured by conventional photon-counting cameras.

Further, we performed a quantitative comparison of the best-possible precision ($\sqrt{\text{CRB}}$) achievable by various imaging techniques under practical imaging scenarios. When molecules are confined within a thin sample, raPol exhibits the best overall precision for measuring in-plane orientation and 2D position simultaneously compared to other methods (\cref{fig:8}(b)). Interestingly, we find that one must defocus the sample slightly (by \textasciitilde200~nm) to obtain the best measurement precision for measuring lateral position and 3D orientation; in this configuration, raPol also outperforms all other methods (\cref{fig:8}(c)). For thick samples, CHIDO exhibits the most uniform precision overall for measuring the 3D orientation and 3D position of SMs, whereas raPol achieves the best peak precision (\cref{fig:9}). CHIDO is also more suitable when supercritical light is undetected since it does not suffer from degraded performance near the focal plane (\cref{fig:3}). Further, since the sizes of the raPol and CHIDO (with a stress coefficient of $0.5\pi$) PSFs are typically within 1-3 times of that of $xy$Pol, it is possible to avoid PSF overlap during imaging by using only slightly smaller emitter densities. We summarize the optimal method for each imaging scenario in \cref{table:optimal_method}.

\begin{table}[h]
\centering
\caption {\textbf{Choosing an optimal method for SMOLM}}
\renewcommand{\arraystretch}{1}
\begin{tabular}{@{}ccc@{}}
\toprule
\multicolumn{2}{c}{thin planar sample} &
  thick 3D sample \\ \midrule
\begin{tabular}[c]{@{}c@{}}in-plane\\ orientation\end{tabular} &
  3D orientation &
  \multirow{2}{*}{\begin{tabular}[c]{@{}c@{}}CHIDO (average \\ precision comparable \\ to raPol, more \\ uniform performance) \end{tabular}} \\ \cmidrule(r){1-2}
\begin{tabular}[c]{@{}c@{}}raPol:\\ in focus\end{tabular} &
  \begin{tabular}[c]{@{}c@{}}raPol: nominal focal\\ plane 200~nm\\ above the sample\end{tabular} &
   \\ \bottomrule
\end{tabular}
\label{table:optimal_method}
\end{table}

Our work provides the first comprehensive comparison of various popular and state-of-the-art SMOLM methods for measuring molecular orientation and position simultaneously. The results show that imaging techniques should be carefully chosen based on the scientific target and the parameters of interest; that is, methods that perform well for measuring a subset of orientation and position parameters do not generally work well for all parameters. Our analysis also shows that early approaches to partition the BFP into multiple linear phase ramps (e.g., the bisected \cite{backer2014bisected} and Tri-spot \cite{Zhang2015} PSFs) perform suboptimally compared to simple defocusing and polarization modulation (e.g., vortex waveplates \cite{zhang2020quantum} and stressed-engineered optics \cite{curcio2019birefringent}). However, these simpler schemes may be less robust to artifacts stemming from motion blur, i.e., translational diffusion; subtle changes in the raPol, $xy$Pol, and CHIDO PSFs due to dipole rotation may be difficult to distinguish from translational motion. The magnitude of estimation error depends on both the PSF \cite{Lu2020} and estimation algorithm in use. Finally, while all of these methods rely on precise phase and/or polarizing optics for their implementation, they all have exhibited good robustness to the broad emission spectra typical of fluorescent molecules in various experiments \cite{Ding2020,backlund2016removing,Lu2020,curcio2019birefringent}.

These observations suggest that developing new polarization detection schemes and polarizing optics could be very beneficial for advancing SMOLM performance. Potential directions include using multiple beamsplitters to construct more detection channels \cite{lippert2017angular}, using different phase masks and PSFs over a sequence of camera exposures, adding interferometric detection \cite{zhang2020quantum}, and combining any of the aforementioned detection strategies with varying illumination polarization states \cite{Forkey2003,Beausang2013,Ha1998,Backer2019,location-orientation_1,Sosa2001,Ha1996,backer2016enhanced}.

\section*{Funding}
National Science Foundation (NSF) (1653777).

\section*{Acknowledgments}
We acknowledge the helpful discussions provided by Jin Lu, Tianben Ding, Tingting Wu, and Hesam Mazidi.
\section*{Disclosures}
The authors declare no conflicts of interest.

\bibliography{references}

\end{document}